\documentclass[aps,twocolumn,superscriptaddress,showpacs,prl]{revtex4}
\usepackage{bm}
\usepackage{amssymb}
\usepackage{graphicx,epsfig}
\usepackage{subfigure}

\begin{document}
\title{Macroscopic Greenberger-Horne-Zeilinger and W States
  in  Flux Qubits}
\author{Mun Dae \surname{Kim}}
\email{mdkim@kias.re.kr}
\affiliation{Korea Institute for Advanced Study, Seoul 130-722, Korea}
\author{Sam Young \surname{Cho}}
\email{sycho@cqu.edu.cn}
\affiliation{  Center for Modern Physics and
 Department of Physics, Chongqing University,
 Chongqing 400044, China}
\affiliation{Department of Physics, The University of Queensland,
Brisbane 4072, Australia}
\date{\today}

\begin{abstract}
We investigate two types of genuine three-qubit entanglement, known
as the Greenberger-Horne-Zeilinger(GHZ) and W states, in a
macroscopic quantum system. Superconducting flux qubits are
considered theoretically in order to generate such states. A phase
coupling is proposed to offer enough strength of interactions between
qubits. While an excited state can be the W state, the GHZ state is formed
at the ground state of the three flux qubits. The GHZ
and W states are shown to be  robust  against external flux fluctuations
 for  feasible experimental realizations.
\end{abstract}

\maketitle

{\it Introduction.}$-$Entanglement plays a crucial role in quantum
 information science. Controllable quantum systems such as photons,
 atoms, and ions have provided the opportunities to generate the
 entanglements.  Recent experiments on two qubits have shown the
 existence of entanglement in different types of {\it
 microscopic} systems. Further, multipartite entanglements such as
 the Greenberger-Horne-Zeilinger(GHZ) \cite{Greenberger} and W
 \cite{Zeilinger92,Dur} states have been demonstrated in recent
 experiments of atoms \cite{Rau}, photons and trapped ions
 \cite{Bouwmeester,Roos}. But, in solid-state qubits it has not yet
 been achieved.

 As a {\it macroscopic} quantum system, superconducting qubit
 systems have been investigated intensively in experiments
 because their system parameters can be controlled to
 manipulate quantum states coherently. Indeed,
 the entanglements between two charge \cite{Pashkin}, phase
 \cite{Berkley,Steffen}, and flux qubits \cite{Izmalkov,Plant} have been
 reported. While the timely evolving states in the experiments of
 charge  qubits \cite{Pashkin} exhibit
 a partial entanglement, the excited level ({\it eigenstate}) of
 capacitively coupled two phase qubits \cite{Steffen} shows higher
 fidelity for the entanglement. The experiments in Ref.
 \cite{Izmalkov} show a possibility that two flux qubits can be
 entangled by a macroscopic quantum tunneling between
 two-qubit states, flipping both qubits.
 Actually, the higher fidelity in the capacitively coupled two
 phase qubits is caused by the two-qubit tunneling processes \cite{Steffen}.
 In a very recent study, the two-qubit tunneling process
 was  theoretically shown to play an important role in generating
 the Bell states, maximally entangled, in the ground
 and excited states \cite{KimCho}.

\begin{figure}[t]
\vspace*{4.5cm}
\includegraphics{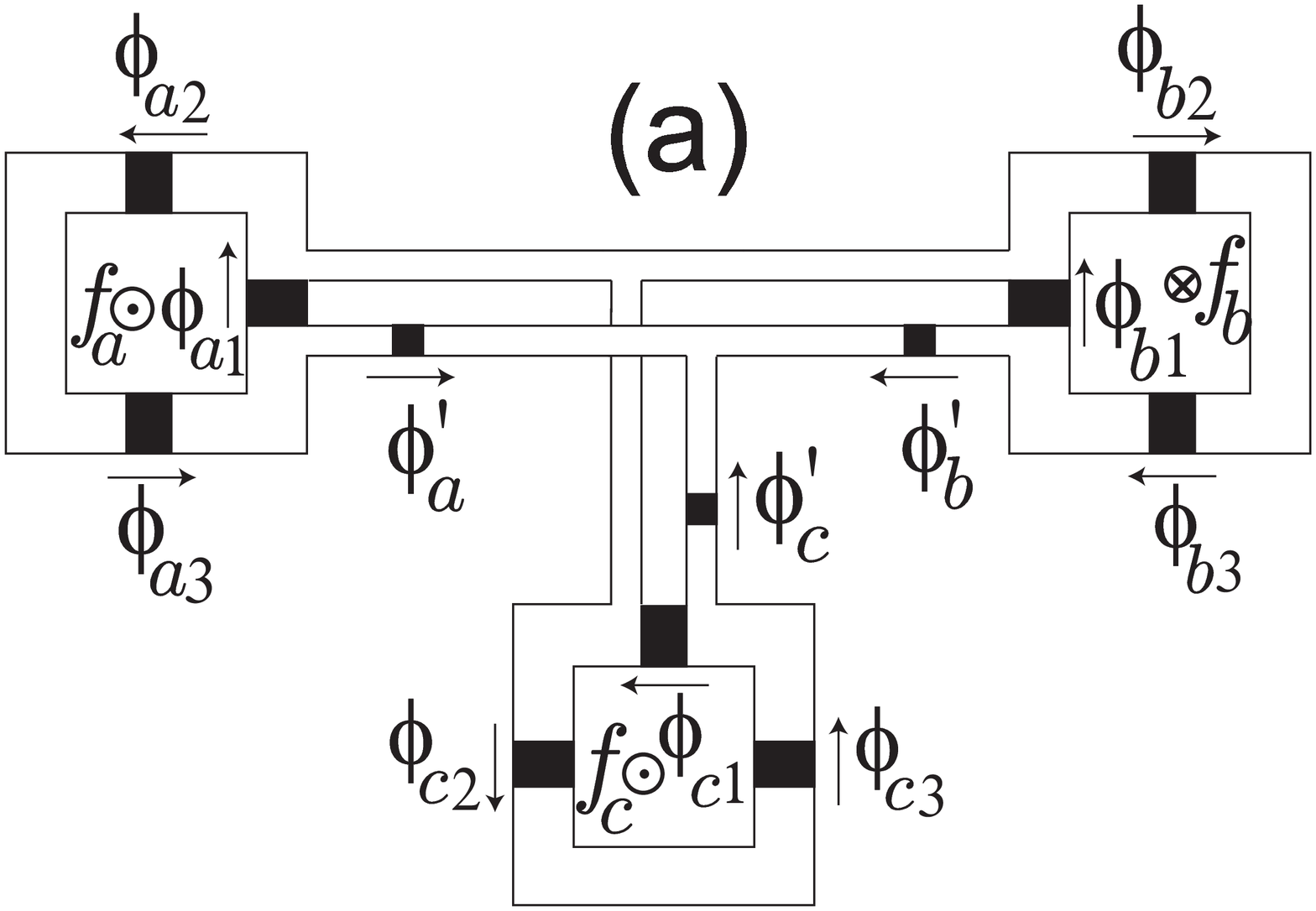}
\vspace*{4.5cm}
\includegraphics{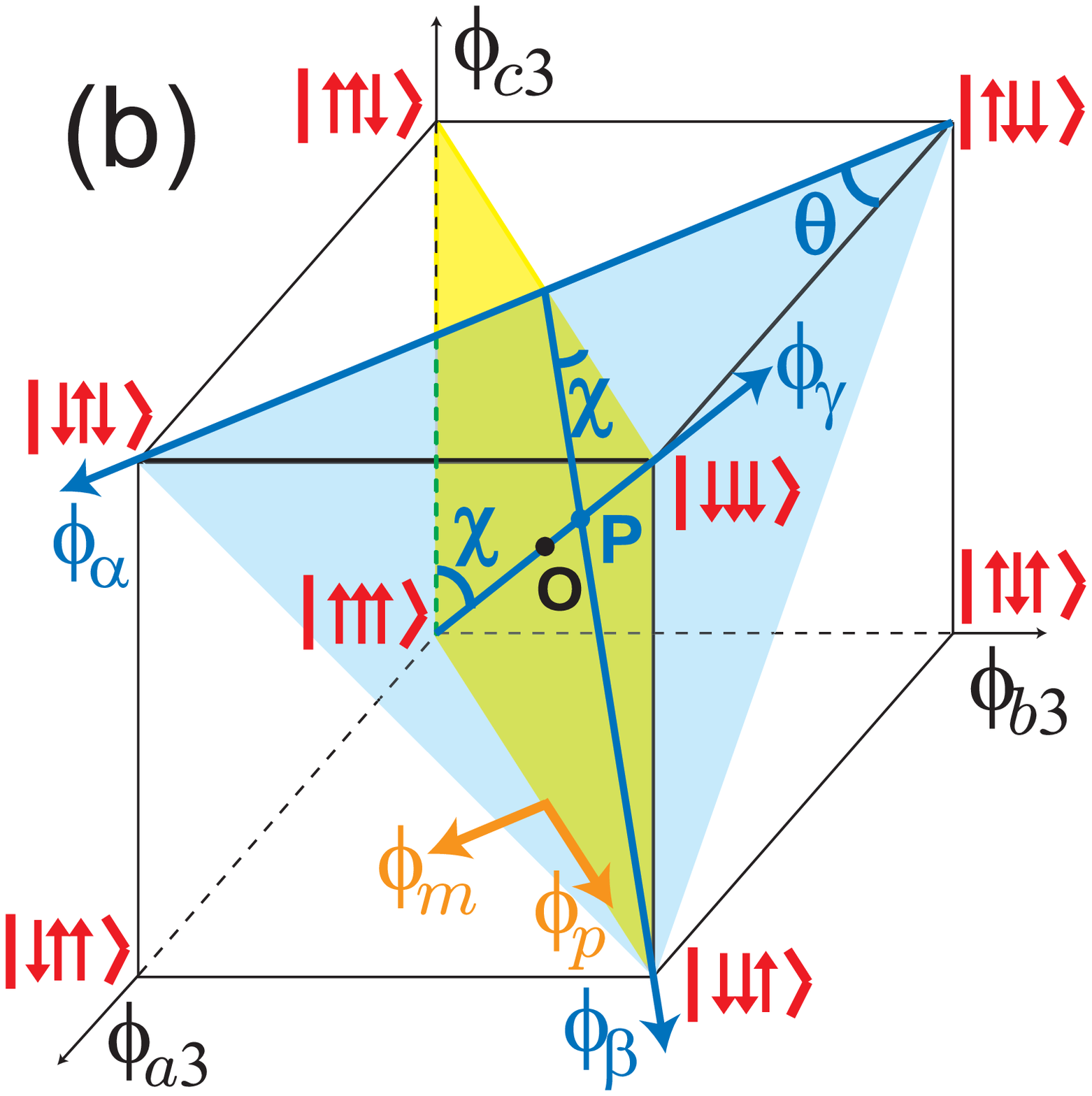}
\includegraphics{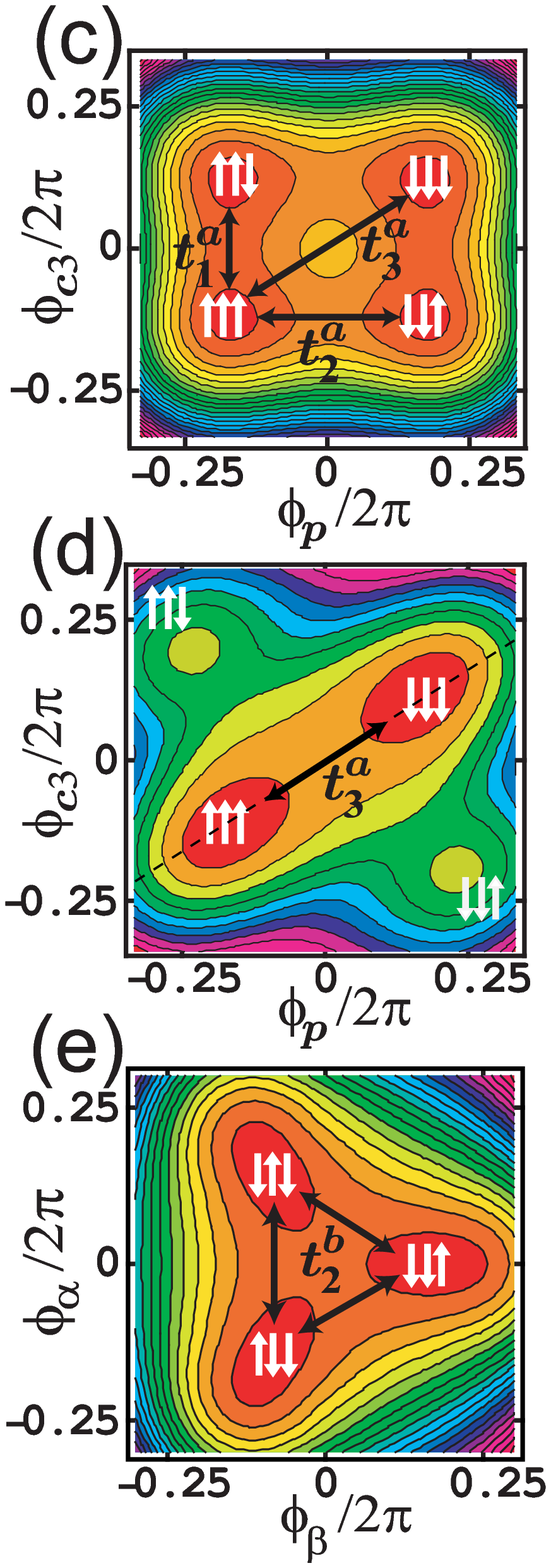}
\vspace{-0.5cm}
\caption{(Color online.) (a) A three flux qubit system.
  The black squares are the Josephson  junctions.
  The Josephson coupling energy of the Josephson junctions
 in the qubit and connecting loop are  $E_{Ji}$ and $E'_J$, respectively.
 $f_{q}$'s  are the external fluxes and
 $\phi$'s are the phase differences  across the junctions.
 (b) The eight states of three qubits are represented
 in $(\phi_{a3},\phi_{b3},\phi_{c3})$-space at the coresonance point.
 $(\phi_\alpha,\phi_\beta,\phi_\gamma)$ are the rotated coordinates and
 O(0,0,0) is the origin of both coordinates. The blue (light gray) triangle intersects
 vertically the $\phi_\gamma$  axis at point P.
 For $E_{J1}/E_J=0.7$, the effective
 potentials $U_{\rm eff}$ in $(\phi_p,\phi_{c3})$-plane ( yellow (dark gray)
 square in (b)) are plotted for (c) $E'_{J}=0$ and (d) $E'_{J}/E_J=0.5$.
 The dotted line in (d)
 coincides with $\phi_\gamma$ axis in (b).
 (e) The effective
 potential $U_{\rm eff}$ in $(\phi_\alpha,\phi_\beta)$-plane (blue (light gray)
 triangle in (b)) for $E'_{J}/E_J=0.05$ and $E_{J1}/E_J=0.75$.
 Here and after, the superscript $a (b)$ in $t^{a (b)}_i$ denotes
 the tunnelling processes including (excluding) the  states,
 $|\uparrow\uparrow\uparrow\rangle$ or  $|\downarrow\downarrow\downarrow\rangle$,
 and $i=1,2,3$ the single-, two-, and three-qubit tunnelling processes, respectively.
}
\label{ThreeQubits}
\end{figure}

 For multipartite entanglements in superconducting qubit systems,
 there have been few studies.
 To produce the GHZ state in three charge qubits,
 only a way of doing a local qubit operation
 via time evolutions  was suggested \cite{Wei}.
 As one of possible directions to produce such multipartite
 entanglements, then, it is natural to ask
 how to create
 the W state as well as the GHZ state
 in the {\it eigenstates} of superconducting three-qubit systems.
%
 Here we consider three flux qubits.
 Normally, the interaction strength between inductively coupled flux qubits
 \cite{Majer} is not so strong  that the controllable range of interaction
 is not sufficiently wide.
 To control a wide range of interaction strengths in the qubits,
we use the phase-coupling scheme \cite{KimTwo, Ploeg, KimCont,Grajcar, Cho07} for
three qubit (see  Fig. \ref{ThreeQubits}(a)) which
 enables  to generate the GHZ and W states
 and  to keep them robust  against external flux fluctuations
 for  feasible experimental realizations.


{\it Model.}$-$We start with the model shown in Fig. \ref{ThreeQubits}(a).
 The Hamiltonian is written by
$\hat{H}=\frac12\hat{ P}^T_i{M}^{-1}_{ij}\hat{ P}_j+U_{\rm eff}(\hat{\bm{\phi}}),$
where $\hat{P}_i =  -i\hbar\partial /\partial \hat{\phi}_i$ and
$M_{ij}=(\Phi_0/2\pi)^2C_i\delta_{ij}$ with the capacitance of the
Josephson junctions  $C_{i}$.
The dynamics of the flux qubits \cite{Mooij} are described by the
phase variables $\hat{\bm{\phi}}=(\phi_{qi}, \phi'_{q})$ with
$q=a,b,c$ and $i=1,2,3$, where $\phi$'s are the phase differences
across the Josephson junctions. If we neglect the small inductive
energy, the effective potential is written in terms of  the
Josephson junction energies, $U_{\rm
eff}(\bm{\phi})=\sum_{q}[\sum^3_{i=1}E_{Ji}(1-\cos\phi_{qi})
+E'_{J}(1-\cos\phi'_q)]$. The periodic boundary conditions
involved in the qubit loops and the connecting loops can be written
as
\begin{eqnarray}
\label{qubit}
\phi_{q1}+\phi_{q2}+\phi_{q3}=2\pi (n_{q}+ f_{q}), \\
\label{connac}
(\phi_{a1}-\phi_{c1})-\phi'_a+\phi'_c=2\pi r, \\
\label{connbc} (\phi_{b1}-\phi_{c1})-\phi'_b+\phi'_c=2\pi s,
\end{eqnarray}
where  $q=a,b,c$ is qubit index and  $r,s,n_q$ integers. Here
$f_{q}\equiv\Phi_{q}/\Phi_0$ with external flux $\Phi_{q}$ and the
superconducting unit flux quantum $\Phi_0=h/2e$. Two independent
conditions in Eqs. (\ref{connac}) and (\ref{connbc}) are the
boundary conditions for connecting loops.
For simplicity  we consider  $E_{J2}=E_{J3}=E_J$
and $C_2=C_3=C$, so we can set $\phi_{q2}=\phi_{q3}$ and Eq.
(\ref{qubit}) becomes $\phi_{q1}=2\pi (n_{q}+ f_{q})-2\phi_{q3}$.
The results for  $E_{J2}\neq E_{J3}$ are qualitatively the same.

At the coresonance point
$(f_a,f_b,f_c)=(1/2,1/2,1/2)$, the  effective potential is given by
\begin{eqnarray}
\label{Ueff}
U_{\rm eff}(\bm{\phi})\!\!&=&\!\!\!\!\!\!\sum_{q=a,b,c}
(E_{J1}\cos2\phi_{q3}-2E_{J}\cos\phi_{q3}-E'_{J}\cos\phi'_q)\nonumber\\
&+&3E_{J1}+6E_J+3E'_J.
\end{eqnarray}
Here, we introduce a rotated coordinates
$\bm{\tilde{\varphi}}=(\phi_\alpha,\phi_\beta,\phi_\gamma)$
in Fig. \ref{ThreeQubits}(b).
The Euler rotations provide new coordinates
such as $\bm{\tilde{\varphi}}^T
={\cal R}_2(\chi,0,0){\cal R}_1(0,0,\theta)\bm{\varphi}^T={\cal R}(\chi,\theta)\bm{\varphi}^T$
with $\chi=-\tan^{-1}\sqrt{2}$, $\theta=-\pi/4$
and $\bm{\varphi}=(\phi_{a3}, \phi_{b3}, \phi_{c3})$,
which can be written explicitly as
\begin{eqnarray}
\label{transform}
\left(
\begin{array}{ll}
\phi_\alpha\\
\phi_\beta\\
\phi_\gamma
\end{array}
\right)
=\frac{1}{\sqrt{6}}
\left(\matrix{\sqrt{3} & -\sqrt{3} & 0 \cr
1 & 1 & -2 \cr
\sqrt{2} & \sqrt{2} & \sqrt{2}}\right)
\left(\matrix{\phi_{a3} \cr \phi_{b3}\cr \phi_{c3}}\right).
\end{eqnarray}
In the same way,  a new coordinates for $\bm{\varphi'}=(\phi'_{a}, \phi'_{b}, \phi'_{c})$
is given as $\bm{\tilde{\varphi}'}^T
={\cal R}(\chi,\theta)\bm{\varphi'}^T$ with  $\bm{\tilde{\varphi}'}=(\phi'_\alpha, \phi'_\beta,
 \phi'_\gamma)$.
Using the boundary conditions of Eqs. (\ref{connac})-(\ref{connbc})
the Hamiltonian  is written in the
transformed coordinates, $\bm{\tilde{\phi}}\equiv
(\phi_\alpha,\phi_\beta,\phi_\gamma,\phi'_\gamma)$, as ${\hat
H}=\sum_{\mu=\alpha,\beta,\gamma} \frac{{\hat
P}_\mu^2}{2M_\mu}+\frac{{\hat P}'^2_\gamma}{2M'_\gamma} +U_{\rm
eff}(\hat{\bm{\tilde{\phi}}}),$ where $M_\alpha=M_\beta=4C_1+2C+2C',
M_\gamma=4C_1+2C$, and $M'_\gamma=C'$. Note that the value of
$\phi'_\gamma$ is determined at the potential minimum, $\partial
U_{\rm eff}(\bm{\tilde{\phi}})/\partial \phi'_\gamma=0$.

 The eight corners of the hexahedron in Fig.
\ref{ThreeQubits}(b) correspond to the three-qubit states.
Here the $|\downarrow \rangle$ $(|\uparrow \rangle)$ is defined as diamagnetic
(paramagnetic) current state which corresponds to positive
(negative) value of $\phi_{qi}$ in the boundary condition of Eq. (\ref{qubit}).
These states can be represented more
clearly in the rotated coordinates, $(\phi_\alpha,\phi_\beta,\phi_\gamma)$,
because the effective potential $U_{\rm eff}(\bm{\phi})$ has three-fold rotational symmetry about the
$\phi_\gamma$-axis, which can be shown as in the follows.
Using the transformation of Eq. (\ref{transform})
one of the terms in Eq. (\ref{Ueff}) is written as
$\sum_{q=a,b,c}\cos\phi_{q3}
=\cos(\phi_\gamma/\sqrt{3})[2\cos(\phi_\beta/\sqrt{6})\cos(\phi_\alpha/\sqrt{2})+\cos(2\phi_\beta/\sqrt{6})]
-\sin(\phi_\gamma/\sqrt{3})[2\sin(\phi_\beta/\sqrt{6})\cos(\phi_\alpha/\sqrt{2})-\sin(2\phi_\beta/\sqrt{6})]$.
Here, if we rotate the potential by $2\pi/3$ about the
$\phi_\gamma$ axis as
$\psi_\alpha=-(1/2)\phi_\alpha-(\sqrt{3}/2)\phi_\beta,
\psi_\beta=(\sqrt{3}/2)\phi_\alpha-(1/2)\phi_\beta$ and
$\psi_\gamma=\phi_\gamma$, we can easily check the invariance of the effective potential,
$U_{\rm{eff}}(\bm{\phi})$.

 In order to study the GHZ state,
$\left|\Psi_{\rm GHZ}\right\rangle=\left(\left|\uparrow\uparrow\uparrow\right\rangle
+\left|\downarrow\downarrow\downarrow\right\rangle\right)/\sqrt{2}$,
we draw the yellow (dark gray) square
introducing the auxiliary coordinates defined by
 $\phi_p\equiv (\phi_{a3}+\phi_{b3})/\sqrt{2}$ and $\phi_m\equiv
(\phi_{a3}-\phi_{b3})/\sqrt{2}$, while for W state,
$\left|\Psi_{\rm W}\right\rangle=\left(
 \left|\uparrow\downarrow\downarrow\right\rangle
 +\left|\downarrow\uparrow\downarrow\right\rangle
 +\left|\downarrow\downarrow\uparrow\right\rangle
 \right)/\sqrt{3}$,
we consider the  blue (light gray) triangle.
Figures \ref{ThreeQubits}(c)-(e) show the
effective potential $U_{\rm eff}(\bm{\phi})$ in Eq. (\ref{Ueff}).
 When the three qubits are decoupled
for $E'_{J}=0$ \cite{KimTwo,KimCont}, Fig. \ref{ThreeQubits}(c)
shows that the single-qubit tunneling, $t^a_1$, is dominant
over the three-qubit tunneling, $t^a_3$. As $E'_J$
increases, it is shown in Fig. \ref{ThreeQubits}(d) that the
three-qubit tunneling becomes dominant.
Then the GHZ state is expected to be formed at the ground state.
The dotted line in Fig. \ref{ThreeQubits}(d)
coincides with $\phi_\gamma$ axis in Fig. \ref{ThreeQubits}(b).
Along the  $\phi_\gamma$ axis, the double-well potential is given by
$U_{\rm eff}(0,0,\phi_\gamma,\sqrt{3}\pi)=
3E_{J1}\left(1+\cos\frac{2\phi_\gamma}3\right)+6E_J\left(1-\cos\frac{\phi_\gamma}3\right),$
where the barrier hight is proportional to $E_{J1}$.
The WKB approximation  allows us to calculate the
three-qubit tunneling, $t^{a}_{3}$, through this double-well
potential \cite{Orlando,KimOne}. Other tunnelings such as single-qubit tunnelings,
$t^a_1$ and $t^b_1$, and two-qubit tunnelings, $t^a_2$ and $t^b_2$, can
also be calculated.
The tight-binding approximation based on the eight states of
three qubits gives the effective Hamiltonian,
$H=\sum_{\nu}E_{\nu}|\nu\rangle\langle \nu|
-\sum_{\nu,\nu'}t_{\nu\nu'}|\nu\rangle\langle \nu'|,$
where
$t_{\nu\nu'}= t^{a(b)}_i$ and $|\nu\rangle=|s_as_bs_c\rangle$ with $s_q \in\{\uparrow, \downarrow\}$.

{\it Q-measure.}$-$
The global entanglement for tripartite systems
can be quantified by the Q-measure \cite{Meyer}.
For a normalized
arbitrary three-qubit state, $|\Psi\rangle=
c_1\left|\downarrow\downarrow\downarrow\right\rangle
+c_2\left|\downarrow\downarrow\uparrow\right\rangle
+c_3\left|\downarrow\uparrow\downarrow\right\rangle
+c_4\left|\downarrow\uparrow\uparrow\right\rangle
+c_5\left|\uparrow\downarrow\downarrow\right\rangle
+c_6\left|\uparrow\downarrow\uparrow\right\rangle
+c_7\left|\uparrow\uparrow\downarrow\right\rangle
+c_8\left|\uparrow\uparrow\uparrow\right\rangle$, the Q-factor is
given by
\begin{eqnarray}
Q(|\Psi\rangle)=\frac43\sum^3_{j=1}D_j(|\Psi\rangle),
\end{eqnarray}
where
$D_1(|\Psi\rangle)=|c_1c_6 - c_2c_5|^2 + |c_1c_7 - c_3c_5|^2
+ |c_1c_8 - c_4c_5|^2 + |c_2c_7 - c_3c_6|^2 + |c_2c_8 -
        c_4c_6|^2 + |c_3c_8 - c_4c_7|^2$ and
$D_2(|\Psi\rangle)$ and $D_3(|\Psi\rangle)$ are obtained by
exchanging the indices  as $3\leftrightarrow 5, 4\leftrightarrow 6$
for $D_2$ and $2\leftrightarrow 5,  4\leftrightarrow 7$ for $D_3$.
For the GHZ state, $Q(|\Psi_{\rm GHZ}\rangle)=1$ and for the W state
$Q(|\Psi_{\rm W}\rangle)=8/9$.

\begin{figure}[t]
\vspace{5.8cm}
\includegraphics{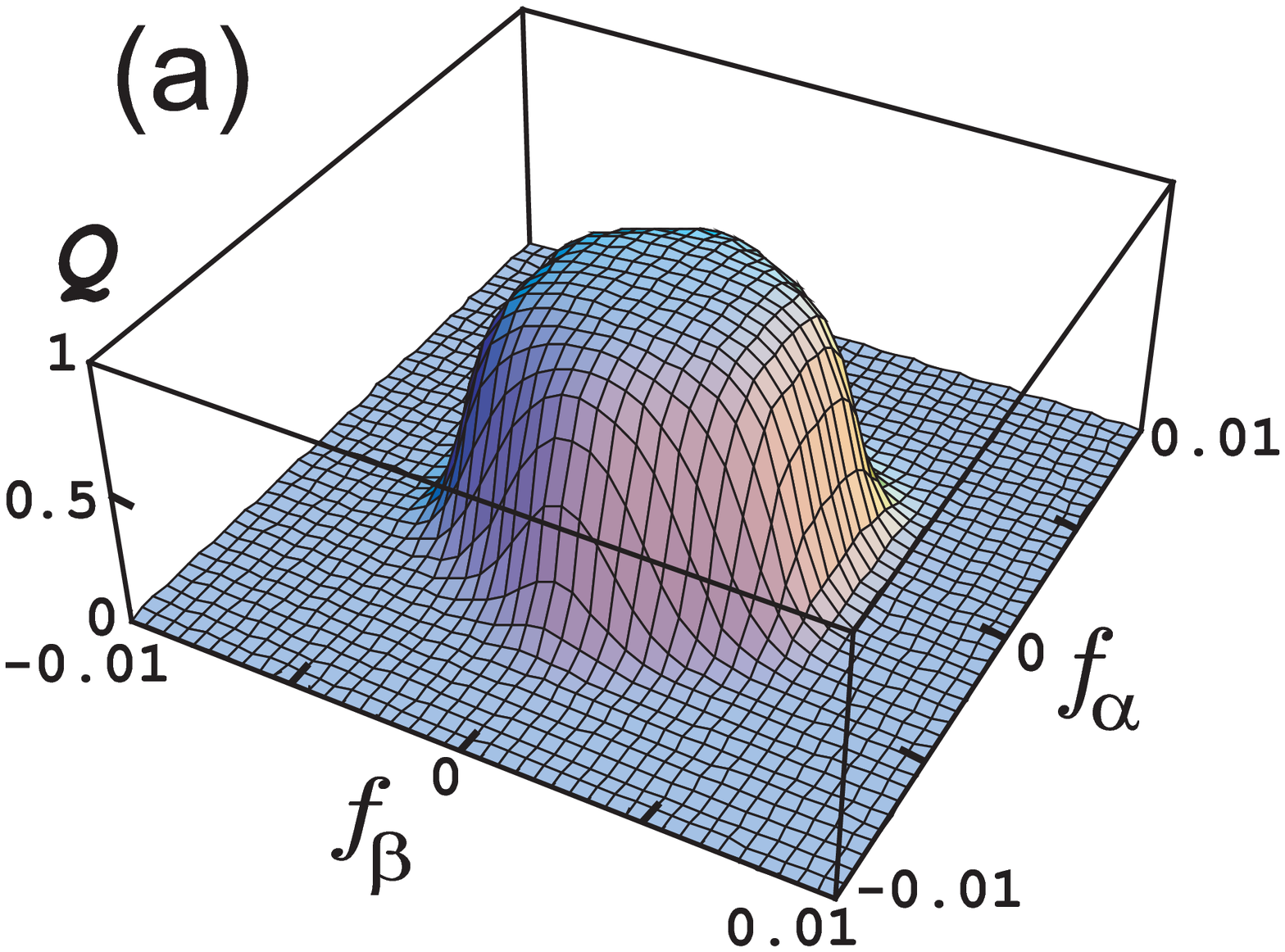}
\includegraphics{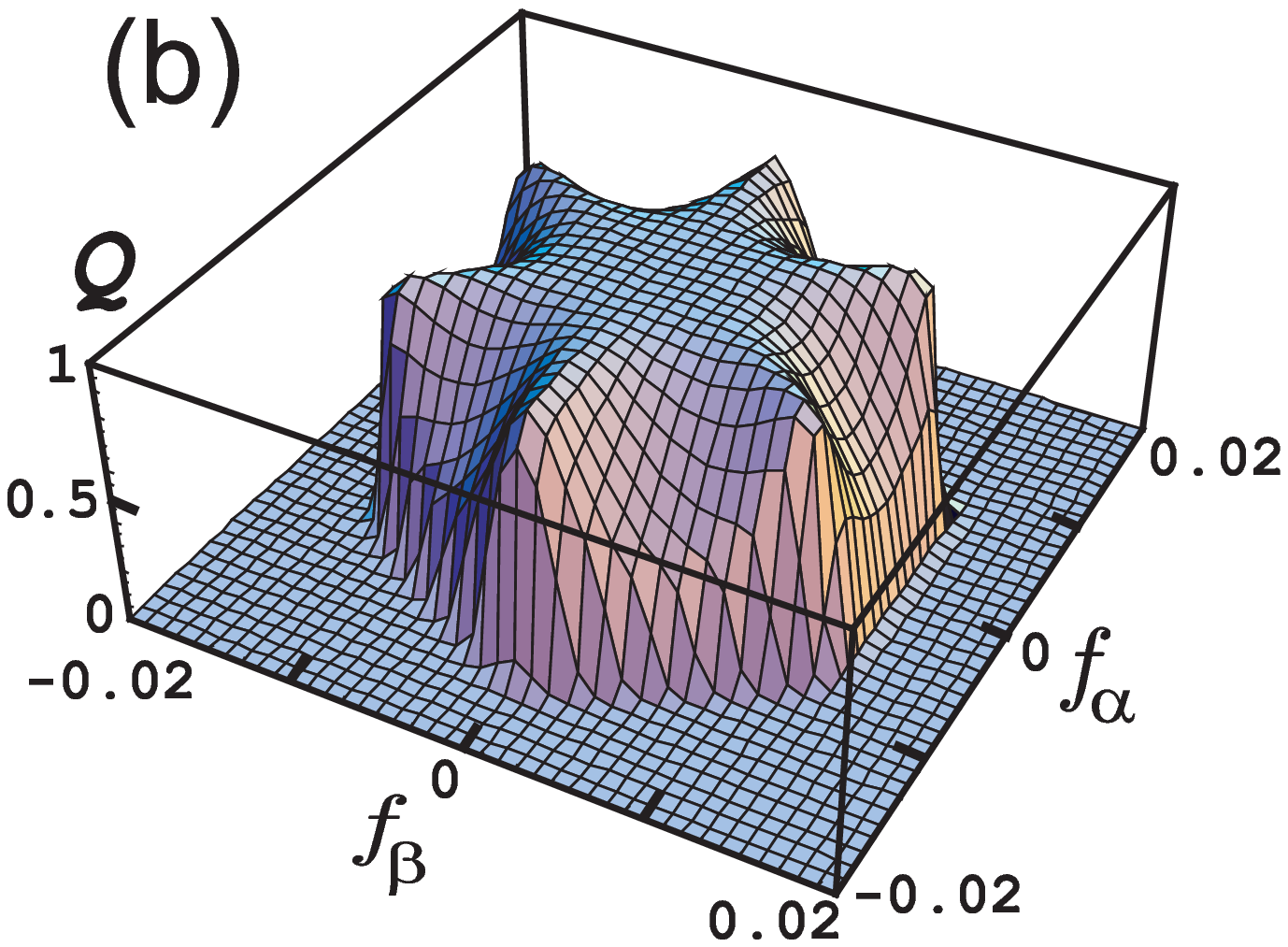}
\vspace{0cm}
\includegraphics{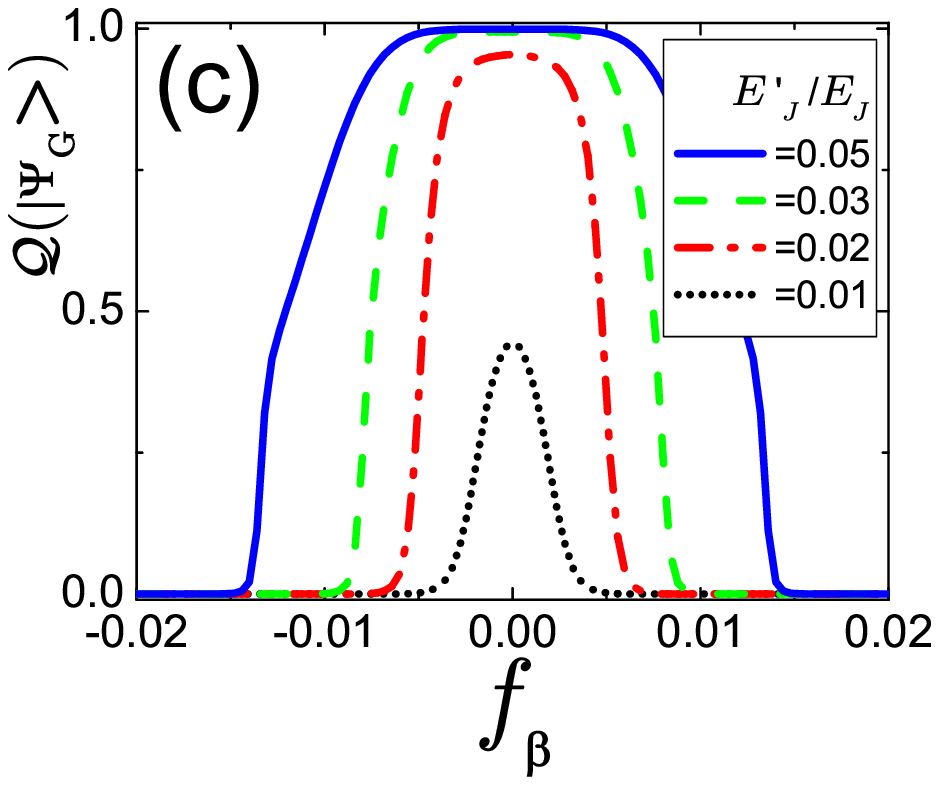}
\hspace{0cm}
\includegraphics{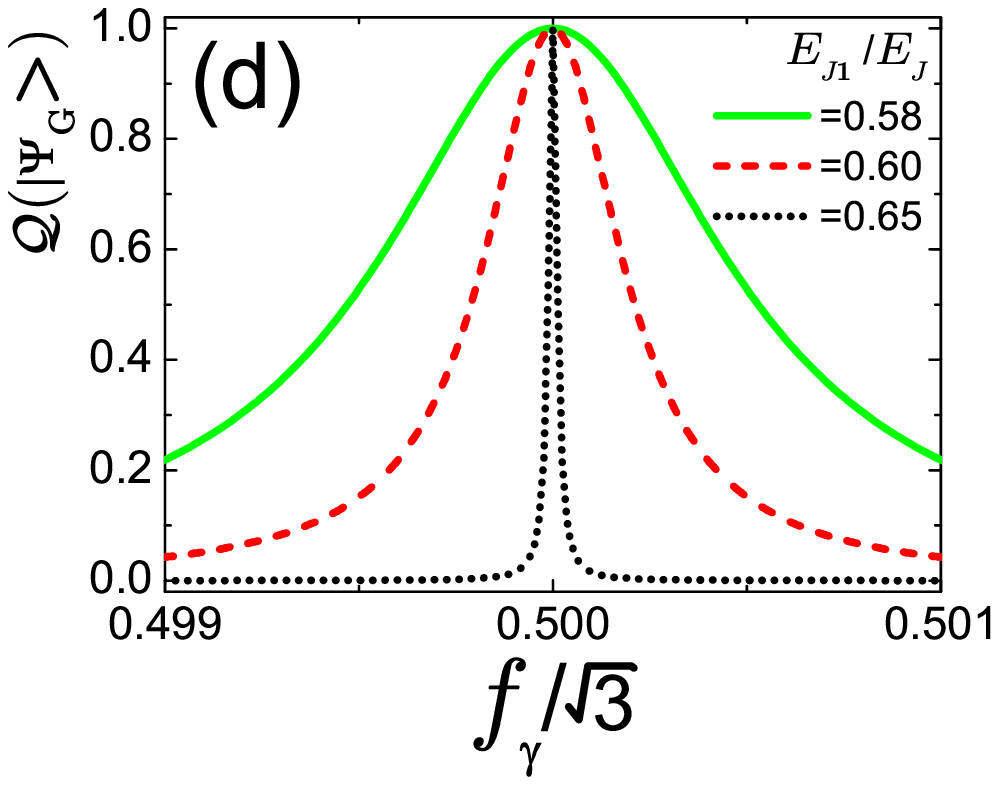}
\vspace*{0.5cm}
\caption{(Color online.) The Q-factors of the ground state
 in the three qubit system
for (a) $E'_J/E_J=0.02$ and for (b) $E'_J/E_J=0.05$. Here,
$f_\gamma = \sqrt{3}/2$ and $E_{J1}/E_J=0.7$.  (c) Cut view of
Q-factors in (a) and (b)   for $f_\alpha=0$. (d) For
$f_\alpha=f_\beta=0$ and $E'_J/E_J=0.6$, Q-factors are plotted as
a function of $f_\gamma$ for several $E_{J1}$.
} \label{GHZQ}
\end{figure}

{\it GHZ state.}$-$We plot the Q-factors for the ground state in
Figs. \ref{GHZQ}(a) and (b) as a function of the rotated fluxes
$f_\alpha \equiv (f_{a}-f_{b})/\sqrt{2},
f_\beta \equiv (f_{a}+f_{b}-2f_{c})/\sqrt{6},
f_\gamma \equiv (f_{a}+f_{b}+f_{c})/\sqrt{3}$.
Note that the coresonance point
$(f_a,f_b,f_c)=(1/2,1/2,1/2)$ is transformed to
$(f_\alpha,f_\beta,f_\gamma)=(0,0,\sqrt{3}/2)$.
For $E'_J/E_J=0.02$ in Fig. \ref{GHZQ} (a), $Q < 1$. But, as $E'_J$
increases, the GHZ state appears around the coresonance point in
Fig. \ref{GHZQ} (b). Figure \ref{GHZQ} (c) is the cut view of
Q-factor for various coupling strength. It is found that for the
GHZ state, $E'_J$ should be larger than $0.03 E_J$. It turns out
that the coupling strength from the inductive coupling scheme
corresponds to $E'_J \approx 0.005E_J$ \cite{KimTwo}. This
shows that the inductive coupling scheme cannot provide a
sufficient coupling for the GHZ state.

To be observed experimentally, the GHZ state should be robust
against fluctuations of external flux. Figure \ref{GHZQ}(b) shows
that the GHZ state can be obtained for a broad range of $f_\alpha$
and $f_\beta$. Thus, let us examine the behavior of Q-factor as a
function of $f_\gamma$ (Fig. \ref{GHZQ}(d)).
If the peak width is too narrow compared with the fluctuations of external flux,
the GHZ state cannot be observed experimentally.
Actually, it is found that the
three-qubit tunneling $t^a_3$ plays an important role for wide
peak width. If other tunneling processes  except $t^a_3$ are negligible,
small flux fluctuations  can influence
the Q-factor given approximately by $Q(|\Psi\rangle)\approx
(t^a_3)^2/((\delta E)^2 + (t^a_3)^2)$, where $\delta E$ is the
energy level change with $E_{\downarrow\downarrow\downarrow}=E_g+\delta E$
and $E_{\uparrow\uparrow\uparrow}=E_g-\delta E$,
$E_g$ the ground state energy
and $\delta E \propto (f_\gamma/\sqrt{3}-0.5)$ \cite{KimTwo}.
Qualitatively, then, $t^a_3$ corresponds to the peak width of the
envelope of Q-factor in Fig. \ref{GHZQ}(d). Consequently, the
stronger $t^a_3$, the wider the range for the GHZ state.

But the single-qubit tunnelling $t^a_1$ makes the coupled qubits unentangled.
Thus we need to suppress the single-qubit tunnelling, while
enhancing the three-qubit tunnelling. In order to do so we need strong
coupling as shown in Fig. 1(d), where the single-qubit tunnellings
between the ground and excited levels become suppressed.
On the other hand the relevant parameter for
$t^a_3$ is the Josephson coupling energy $E_{J1}$.
As  $E_{J1}$ decreases, the barrier
in the double-well potential in Fig. \ref{ThreeQubits}(d) becomes lower.
It implies that, to get a larger value of $t^a_3$, $E_{J1}$ should be
smaller. But too small $E_{J1}$ makes some excited states, $|s_as_bs_c \rangle$, unstable.
We show the minimum $E_{J1}$'s in Table I for three representative $E'_J$'s.
In fact, we found that for strong coupling case the excited states
are stable for smaller $E_{J1}$. For small $E'_J/E_J=0.05$ we
obtained $t^a_1/E_J=1.3 \times 10^{-3}$ and thus $t^a_3/t^a_1 \approx 5.4 \times 10^{-3}$.
But for larger value of $E'_J/E_J=0.6$ we obtained $t^a_3/t^a_1 \approx 6.4$ with
$t^a_1/E_J=8.0 \times 10^{-4}$. Hence for strong  coupling case
we can expect higher Q-factor for the GHZ state.

In Table I  the peak widths for both GHZ and W states
are calculated at 95\% of the maximum value of Q-factor,
which are approximately proportional to $t^a_3$ and $t^b_2$, respectively.
During the Rabi oscillations the fluctuation of flux is estimated to be
in the order of $10^{-6} [\Phi_0/Hz^{1/2}]$ \cite{Bertet} and
$1/f$ critical current fluctuations of the Josephson junctions is rather weak.
In recent experiments for flux qubits, the flux amplitudes are controlled
up to the accuracy of $10^{-5}\Phi_0$. In this respect, the peak width,
$\delta f_\gamma \sim 4\times 10^{-4}$, for $E'_J=0.6E_J$ will be sufficient
to observe the GHZ state experimentally.

\begin{table}[b]
\begin{center}
\begin{tabular}{c|c|c|c|c|c}
\hline
\hline
$E'_J$ & $E_{J1}$  & $t^a_3$ & peak width & $t^b_2$ & peak width \\
 &   & & GHZ ($\delta f_\gamma$) & &  W ($\delta f_\alpha$)\\
\hline
0.05 & 0.7 & 7.0 $\times 10^{-6}$ & $\sim\!\!5\times 10^{-7}$ & 6.3 $\times 10^{-4}$ & $\sim\!\! 10^{-4}$  
\\
0.1 & 0.75  & 2.6 $\times 10^{-7}$ & $\sim\!\!2\times 10^{-8}$ & 1.0 $\times 10^{-4}$ & $\sim\!\! 2\times 10^{-5}$ 
\\
0.6 & 0.58 &  5.1 $\times 10^{-3}$ & $\sim\!\! 4\times 10^{-4}$ & 0 & 0  
\\
\hline \hline
\end{tabular}
\end{center}
\caption{Peak widths for Q-factors of GHZ state in Fig. \ref{GHZQ}(d) and
of W state in Fig. \ref{WQ}(d) at 95\% of the maximum values.
Here the unit of $E'_J$, $E_{J1}$, and $t$ is $E_J$.}
\label{table}
\end{table}

 {\it W state.}$-$
 In Fig. \ref{ThreeQubits}(b),
 we present the  blue (light gray) triangle whose corners correspond
 to the three states consisting of the W state,
 $|\Psi_{\rm W}\rangle=\left(
 \left|\uparrow\downarrow\downarrow\right\rangle
 +\left|\downarrow\uparrow\downarrow\right\rangle
 +\left|\downarrow\downarrow\uparrow\right\rangle
 \right)/\sqrt{3}$.
The  blue (light gray) triangle intersects $\phi_\gamma$-axis  at $\phi_\gamma > 0$.
 Actually, there is another intersection plane
 with $\phi_\gamma <0$ for another possible W state.
 For simplicity, we will focus on the W state on the  blue (light gray) triangle  plane.
 The effective potential at the plane
 of the  blue (light gray) triangle for the three states
 is drawn in  Fig. \ref{ThreeQubits}(e).
 Energetically, in our model,
 the energies of three states are higher than those of the two states
 consisting of the GHZ state, i.e., the ground state.
 Then, the W state can be observed in an excited state.

\begin{figure}[t]
\vspace*{7.3cm}
\includegraphics{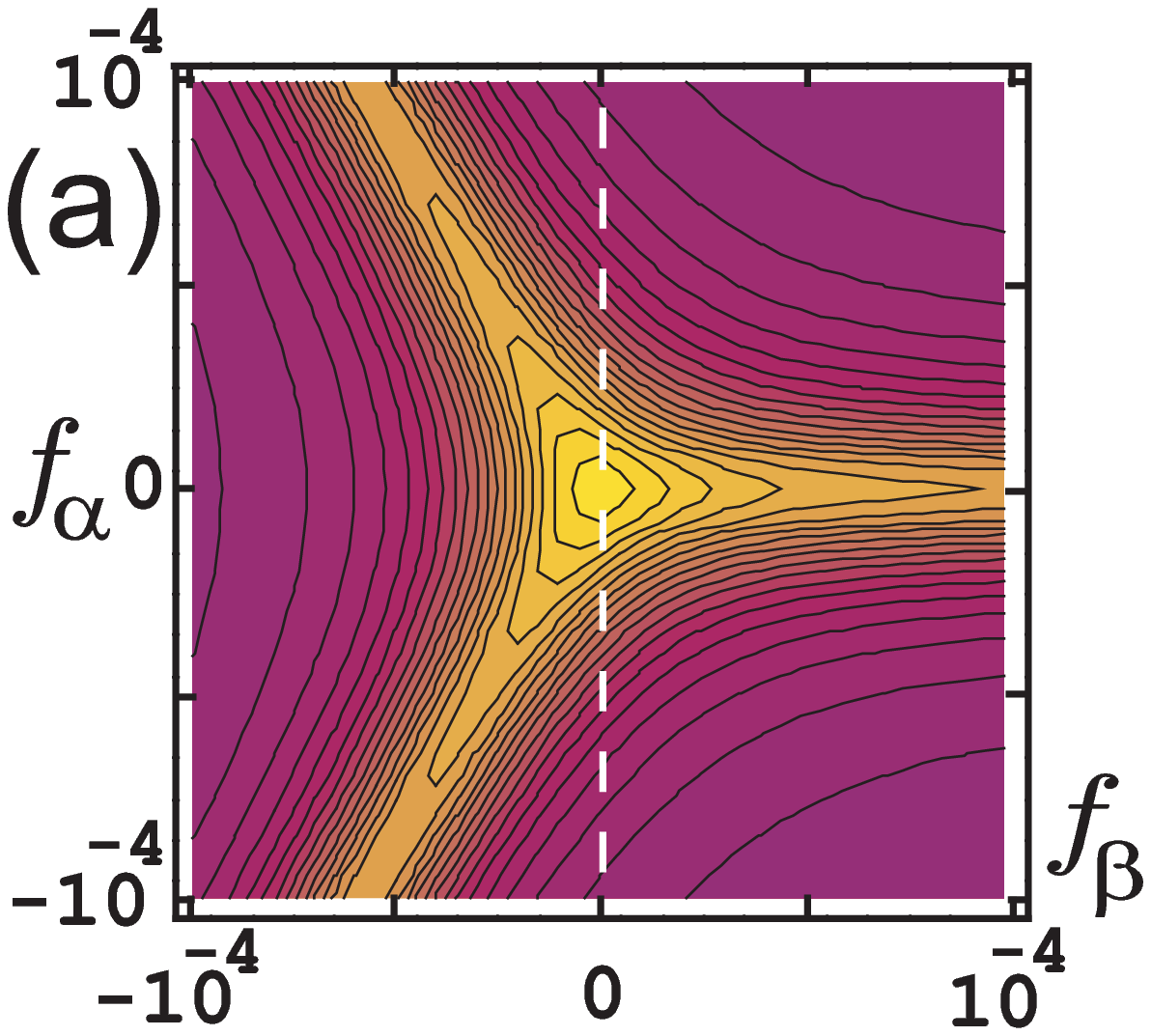}
\includegraphics{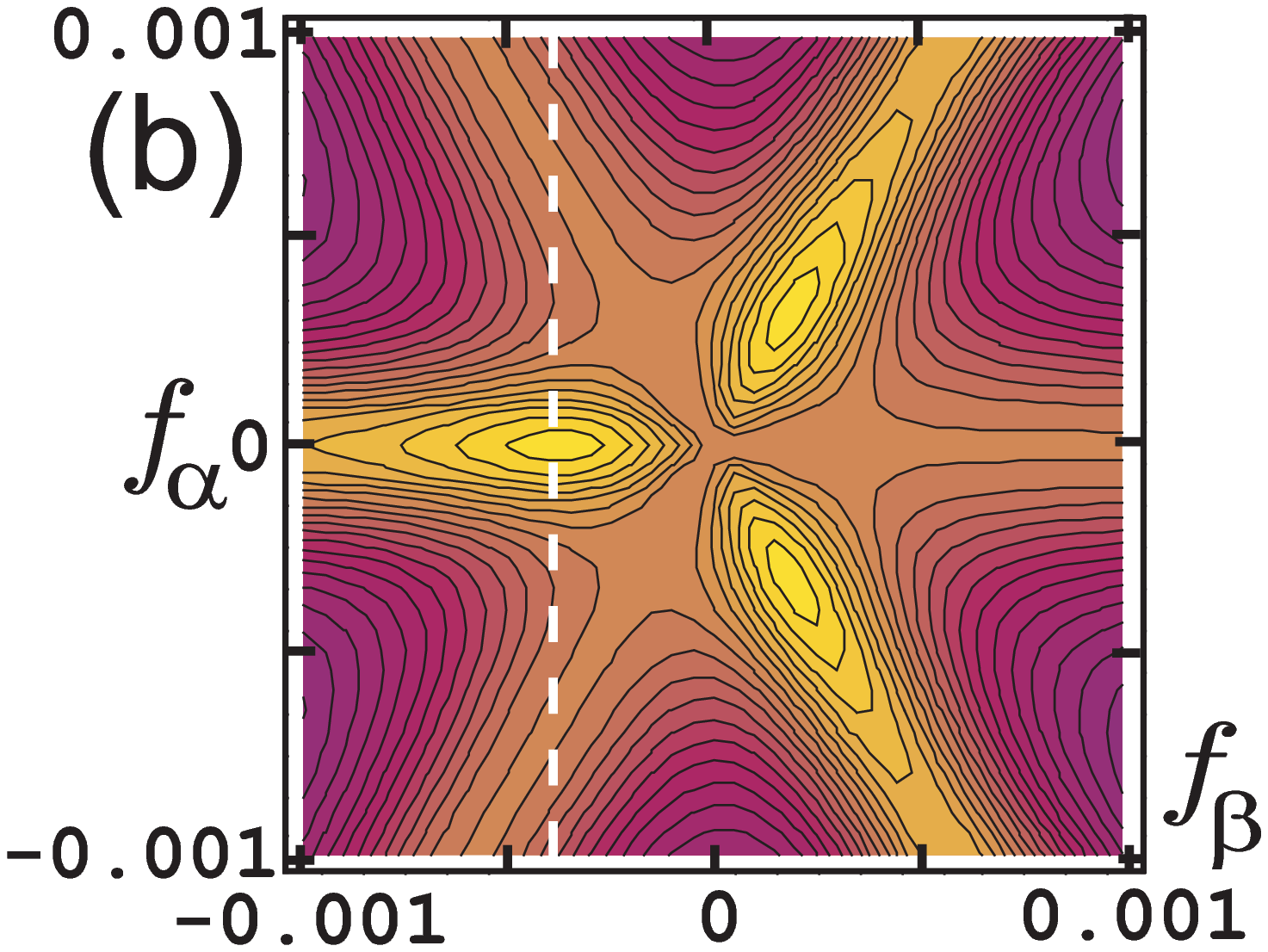}
\includegraphics{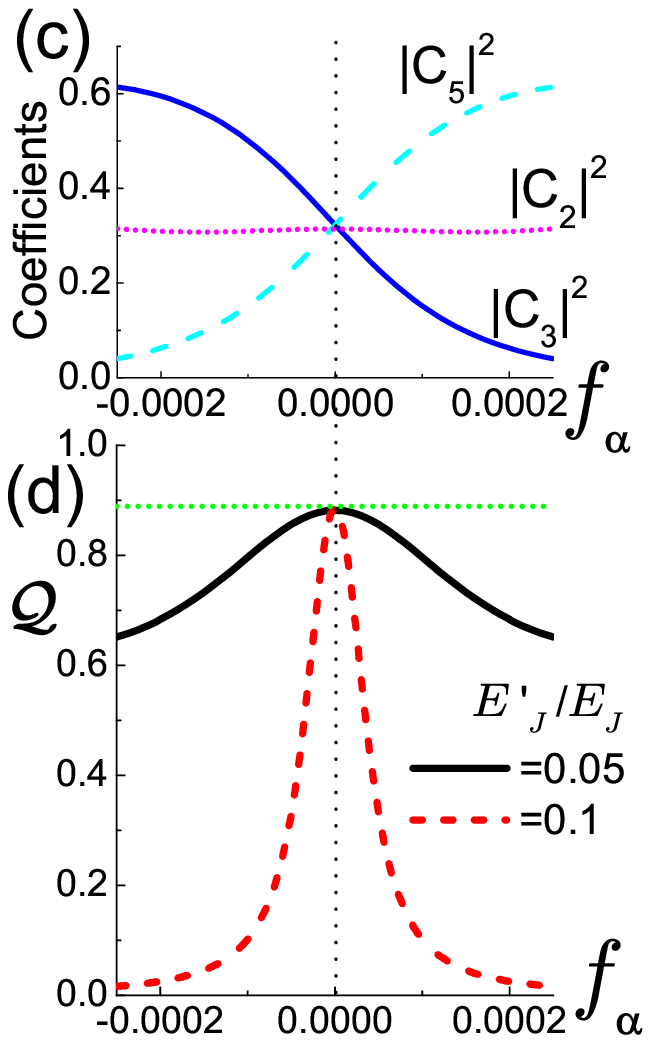}
\vspace{-0.5cm}
\caption{(Color online.)
 The Q-factors of the second excited state with
 $f_\gamma = \sqrt{3}/2-0.01$ for
(a) $E'_J/E_J=0.1$ and $E_{J1}/E_J=0.75$ and (b) $E'_J/E_J=0.05$
and $E_{J1}/E_J=0.7$. Note the different scales and positions of
maximum Q-factor in both figures. Here
the yellow (light gray) regions denote high Q-factors.
(c) The coefficients of the
eigenstate used in (b) are plotted along the
dotted line with $f_\beta=-0.0004$, which shows that W state
is formed around $f_\alpha=0$. (d) Cut view of Q-factors along
dotted lines in (a) and (b). The green dotted line indicates
$Q=8/9$ for W state. The peak width for $E'_J/E_J=0.05$ is much
wider than that for $E'_J/E_J=0.1$.} \label{WQ}
\end{figure}

 Let us discuss how a W state can be realized in an excited state.
 At the coresonance point
 $(f_\alpha,f_\beta,f_\gamma)=(0,0,\sqrt{3}/2)$, the six states
 except for $\{ \left|\downarrow\downarrow\downarrow\right\rangle,
 \left|\uparrow\uparrow\uparrow\right\rangle \}$ are degenerated in
 the second excited state.
 The six states are classified into two classes,
 $\{\left|\uparrow\uparrow\downarrow\right\rangle,
    \left|\uparrow\downarrow\uparrow\right\rangle,
    \left|\downarrow\uparrow\uparrow\right\rangle\}$
    with $S_z=1/2$
 and
 $\{ \left|\uparrow\downarrow\downarrow\right\rangle,
     \left|\downarrow\uparrow\downarrow\right\rangle,
     \left|\downarrow\downarrow\uparrow\right\rangle \}$
     with $S_z=-1/2$.
 Hence, the two classes of the six states can be separated by
 applying an additional flux $\Delta f_\gamma$.
 The three states of each class can form a W state.
 As shown in Fig. \ref{ThreeQubits}(e),
 the two-qubit tunneling amplitude $t^b_2$
creates the W-state, while  the single-qubit tunneling $t^b_1$ destroys the W-state
 because it induces a superposition  of states of the two classes.
 If other small tunnelings are negligible, then,
 the Q-factor is given by
  $Q(|\Psi\rangle)\simeq 8(1 +
 2.5(t^b_1/\Delta E)^2)/9(1 + 4(t^b_1/\Delta E)^2)$.
 From $Q(|\Psi_{\rm W}\rangle)=8/9$,
 it turns out that $|\Delta E| \propto
 |f_\gamma/\sqrt{3}-1/2|$ should be much larger than $t^b_1$.
 Therefore, a sufficient $\Delta f_\gamma$
 are needed to generate a W state.
 Actually, we found that $|\Delta f_\gamma| \gtrsim 0.01$ is sufficient to show
 the generation of a W state  (Figs. \ref{WQ}(a) and (b)).
In Fig. \ref{WQ}(b),  the W state is formed slightly away from
the point, $f_\alpha=f_\beta=0$. For a  relatively weak coupling the
single-qubit tunneling $t^b_1$ as well as $t^b_2$ becomes larger.
Thus, an additional small flux $f_\beta \approx 0.0004$ will break
the symmetry so that the state $|\Psi\rangle$ closer to the W state would be
formed.
 In Fig. \ref{WQ}(c)  we can see the W state
 around $f_\alpha = 0$.

 On the contrary to the GHZ state where the range of $f_\gamma$ is critical
for experimental observation, for W state the range in $(f_\alpha,
f_\beta)$-plane is important as shown in Figs. \ref{WQ}(a) and (b).
Fig. \ref{WQ}(d) shows the Q-factor for W-states whose
peak widths depend on the value of the two-qubit tunneling amplitude $t^b_2$ (Table I).
As $E'_J$ decreases, $t^b_2$ becomes larger. However, if the
coupling strength becomes too weak, the two classes with $S_z=\pm
1/2$ will become overlapped with each other through the single qubit tunnelling
$t^b_1$ so that the W state may
readily be broken. Hence, as a consequence of  compromise,
the W state emerges  for an intermediate coupling strength, $E'_J=0.05E_J$,
with rather broader peak width as shown in Table I.

{\it Discussions and summary.}$-$The quantification of entanglement can be done
by using the state tomography measurement \cite{Steffen,Liu}.
Recently for capacitively coupled phase
qubits the tomography measurement has been done \cite{Steffen},
where they simultaneously measure the state of
coupled qubits. For present coupling we
expect that the similar tomography measurement can also be
performed.

The tripartite entanglement with superconducting qubits has not yet been achieved so far.
For the bipartite entanglement capacitively coupled phase qubits showed high fidelity
in a recent experiment \cite{Steffen}, while for charge qubits only partial entanglement was observed.
The interaction between phase qubits are XY-type interaction which describes simultaneous
two-qubit flipping processes. The two- or multiple-qubit tunnelling processes
are essential for entanglement of qubits \cite{KimCho}.
However, for charge qubits, the interaction are mainly Ising-type.
We believe that this is the reason for weak entanglement in experiments with charge qubits.
For three coupled phase qubits the lowest energy state is $|000\rangle$ state while
the highest is $|111\rangle$ state. Hence the superposition between these two states
will be negligibly weak, thus the GHZ state cannot be formed. But, since the other states,
for example $|100\rangle, |010\rangle, |001\rangle$ states, are energetically degenerated,
the W state could be obtained.

In summary, we investigate a three superconducting flux
qubit system. The GHZ and W states can be realizable in the
eigenstates of the macroscopic quantum system.
We show that while the GHZ state
needs strong coupling strength, the W state can be formed at an
optimized coupling strength.
Moreover, to keep the tripartite
entangled states robust against external flux fluctuations for
feasible experimental realizations,
the three coupled qubit system can provide  relatively
large three-qubit and two-qubit tunneling amplitudes
for GHZ and W states, respectively.

 SYC acknowledges the support from
  the Australian Research Council.

\end{document}